\documentclass[journal]{IEEEtran}
\usepackage{xcolor,soul,framed} 
\usepackage{graphicx}
\usepackage{amsmath}
\usepackage{array}
\usepackage{mdwmath}

\usepackage{mdwtab}
\usepackage{eqparbox}
\usepackage{url}
\usepackage{lineno,hyperref}
\usepackage{amsmath,amssymb}
\usepackage{booktabs}
\usepackage{makecell}
\usepackage{multirow}
\usepackage{multirow}
\usepackage{tabu,bm}
\usepackage{color}
\usepackage{lipsum,mwe,cuted}
\usepackage{stfloats}
\usepackage{cite}
\usepackage{enumitem} 
\usepackage{diagbox}
\usepackage{subcaption}

\title{Semantic Communication Enabled Holographic Video Processing and Transmission}

\author{Jingkai Ying, Zhiyuan Qi, Yulong Feng, Zhijin Qin, Zhu Han, Rahim Tafazolli, and Yonina C. Eldar
\thanks{Jingkai Ying, Zhiyuan Qi, and Zhijin Qin are with the Department of Electronic Engineering, Tsinghua University, Beijing 100084, China, and also with the Beijing National Research Center for Information Science and Technology (BNRist), Beijing 100084, China, and also with the State Key Laboratory of Space Network and Communications, Beijing 100084, China. (e-mail: yjk23@mails.tsinghua.edu.cn; qzy24@mails.tsinghua.edu.cn; qinzhijin@tsinghua.edu.cn). }
\thanks{Yulong Feng is with the State Key Laboratory of Mobile Network and Mobile Multimedia Technology, Shenzhen 518055, China, and also with ZTE Corporation, Shenzhen, 518055, China. (e-mail: feng.yulong1@zte.com.cn). }
\thanks{Zhu Han is with the Department of Electrical and Computer Engineering, University of Houston, Houston, TX 77004 USA, and also with the Department of Computer Science and Engineering, Kyung Hee University, Seoul 446-701, South Korea. (e-mail: hanzhu22@gmail.com).}
\thanks{Rahim Tafazolli is with the Institute for Communication Systems (ICS), 5G/6G Innovation Centre, University of Surrey, GU2 7XH Guildford, UK. (e-mail: r.tafazolli@surrey.ac.uk).}
\thanks{Yonina C. Eldar is with the Faculty of Mathematics and Computer Science, Weizmann Institute of Science, Rehovot 7610001, Israel. (e-mail: yonina.eldar@weizmann.ac.il).}}

\begin{document}

\maketitle

\begin{abstract}
Holographic video communication is considered a paradigm shift in visual communications, becoming increasingly popular for its ability to offer immersive experiences. This article provides an overview of holographic video communication and outlines the requirements of a holographic video communication system. Particularly, following a brief review of semantic communication, an architecture for a semantic-enabled holographic video communication system is presented. Key technologies, including semantic sampling, joint semantic-channel coding, and semantic-aware transmission, are designed based on the proposed architecture. Two related use cases are presented to demonstrate the performance gain of the proposed methods. Finally, potential research topics are discussed to pave the way for the realization of semantic-enabled holographic video communications.
 
\end{abstract}

\section{Introduction}

Holographic video is a revolutionary information modality, which provides panoramic video content and an immersive experience based on three-dimensional view and high-resolution holograms \cite{akyildiz2022holographic}. Holographic video communication (HVC) is regarded as the dominant paradigm for future visual-type communications. It is considered the potential method to realize metaverse and enable numerous applications, such as holographic conferencing, education, and entertainment. However, due to the explosive traffic demand and stringent latency and reliability requirements, it is extremely challenging for today's wireless communication system to support HVCs. 

Sixth-generation (6G) wireless identifies immersive communication as one of its usage scenarios with holographic communication being a typical use case. Despite 6G's anticipated capabilities of providing $1\sim10$ Gbps user experience data rate, under $1$ ms latency, $10^{-5}$ level packet-error rate \cite{wang2023road}, it cannot be guaranteed that 6G will fully support high-quality HVC services. Therefore, more targeted research on architectures and technologies specific to HVC is essential to advance its further development. 

Most existing works on holographic video transmission are in the early exploratory stages, focusing on improving and optimizing specific aspects of the transmission system. An HVC architecture based on existing transmission and networking technologies was proposed in \cite{clemm2020toward}, with emphasis on the transmission of multi-sensory data and the enhancement of current systems through cloud computing and edge computing, as well as software-defined network. The impacts of immersive video transmission protocols on latency and adaptive bitrate transmission were proposed in \cite{van2023tutorial}. Aforementioned work is constrained by the existing communication system, ultimately aiming to establish more informative bit representations and more reliable bit transmission pipelines. However, these methods pile up resources to meet the demanding quality of service requirements for HVC. To facilitate HVC, we pursue a more efficient communication paradigm beyond traditional bit-based schemes.

Thanks to the remarkable advancement of deep learning (DL), semantic communication which is aimed at conveying the meaning of messages, has attracted researchers' attention in recent years \cite{qin2024ai}. Semantic communication can effectively extract and compress meaningful information from source messages, reducing the bandwidth requirements for holographic video transmission by leveraging the powerful nonlinear representation capabilities of DL models. Furthermore, by constructing an end-to-end semantic-channel joint coding architecture and employing joint training, semantic communication can effectively counteract channel impairments and provide globally optimal performance. In \cite{huang2021aitransfer}, a DL-based semantic codec module for HVC, AITransfer, demonstrated its powerful capacity to reduce the burden on transmission. Based on AITransfer, an adaptive control scheme, which can match the model size with the network conditions, was proposed in \cite{zhu2022semantic}. These works demonstrate the effectiveness of semantic communication in HVC and provide preliminary validation of its feasibility for practical deployment. 

It is worth noting that, the ultimate form of holographic communication should encompass the processing and transmission of multi-sensory data, including visual, auditory, olfactory, and tactile information. However, the realization of olfactory and tactile information communication is still difficult, mainly due to the limitation of physical devices. And the auditory information processing is fairly mature. So in this article, we focus solely on the visual aspect which is the primary perceptual source of a holographic experience.

This article introduces semantic communication technologies into HVCs and proposes a semantic enabled HVC solution. The main contributions of this article can be summarized as follows.
\begin{itemize}
\item A novel semantic-aware holographic video communication architecture is proposed.
\item A semantic-aware sampling method based on an attention mechanism is designed, which can capture key areas of point clouds.
\item An efficient and robust joint semantic-channel coding and modulation scheme is developed, which can transmit point clouds adaptively based on semantic features and channel conditions.
\end{itemize}

The remainder of the article is structured as follows. In Section \ref{Section II: Overview of Holographic Video Communications and Semantic Communications}, basic ideas of HVC and semantic communication are introduced. In Section \ref{Section III: Semantic-Aware Holographic Video Communications}, the architecture and key technologies tailored for HVCs are presented. Two related use cases are introduced in Section \ref{Section IV: Use Cases}. Section \ref{Section V: Conclusions and Outlooks} concludes this article with several outlooks.

\section{Overview of Holographic Video Communications and Semantic Communications}
\label{Section II: Overview of Holographic Video Communications and Semantic Communications}

In this section, we start with the definition of holographic video and then summarize its data representation methods and transmission requirements. Subsequently, we analyze the advantages of point clouds and introduce relevant compression methods. Finally, we briefly present the idea of semantic communication.

\subsection{Holographic Video Communications}  
\begin{figure}[t]
\centering
\includegraphics[width=0.95\columnwidth]
{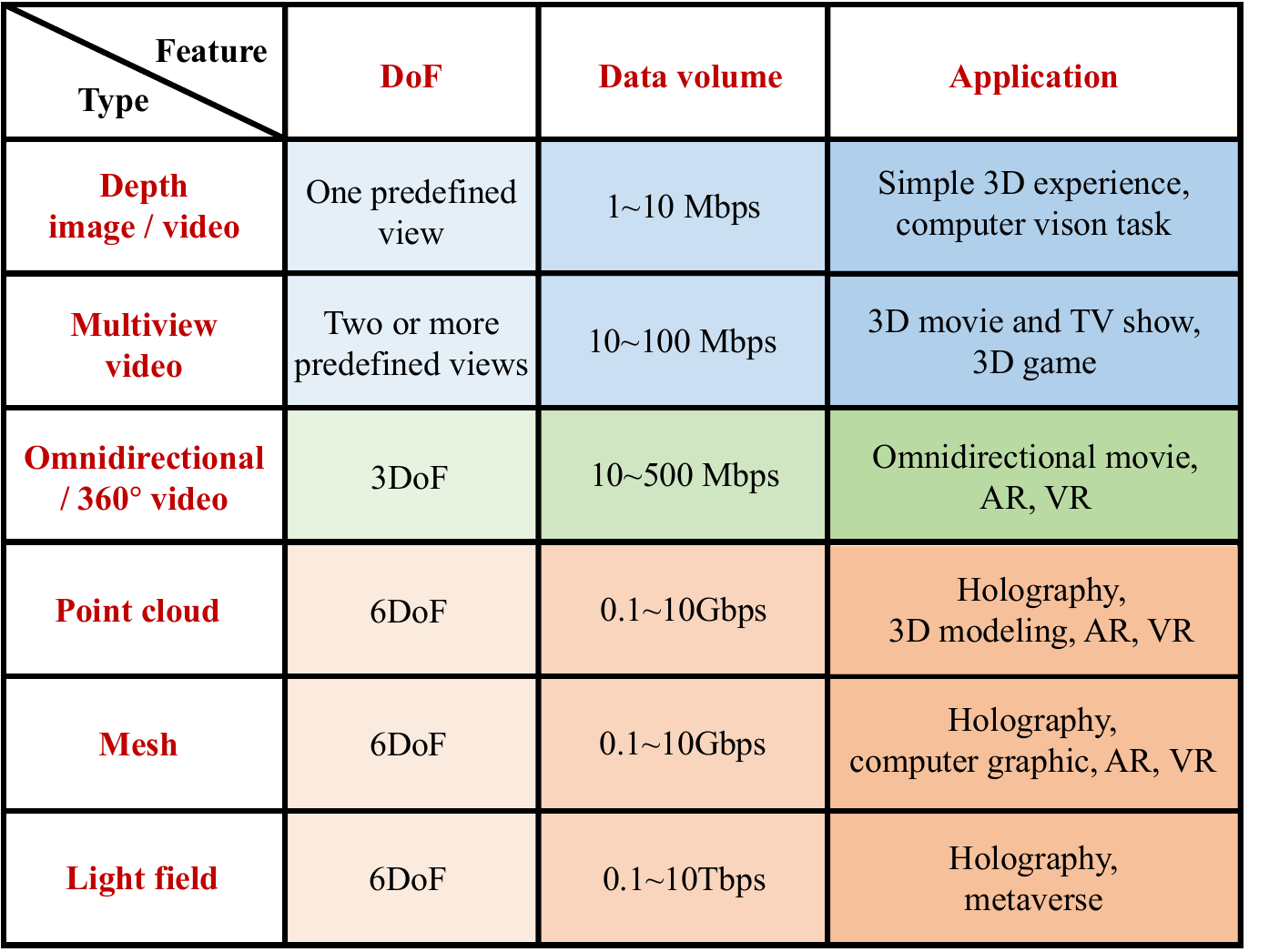}
\caption{Main methods of 3D data representations.}
\label{representations}
\end{figure}

Holographic video is composed of a series of temporally sequential holograms. Ideally, a hologram aims to record all optical information of a three-dimensional (3D) object. The ideal hologram is the ultimate goal of 3D representations. With advancements in hardware, software, and computational technologies, 3D representations are now progressively approaching an ideal hologram. Fig. \ref{representations} illustrates the main types of 3D data representations and their key features. Specifically, depth images/videos are the simplest form of 3D data representation, offering a single-view 3D visual effect. Currently, they are primarily used to enhance the performance of computer vision tasks, like semantic segmentation by leveraging depth information. Multiview video provides multiple predefined viewpoints to represent 3D information. The omnidirectional video ($360^{\circ}$) offers 3 degrees of freedom (3-DoF) visual presentation, including yaw, pitch, and roll, making it suitable for applications like virtual reality (VR) and augmented reality (AR). Point cloud, mesh, and light field can enable 6-DoF, which adds translation movements along each axis based on yaw, pitch, and roll. Point cloud consists of 3D coordinates and additional attributes, which allows for efficient and precise 3D modeling. Composed of more complex vertices, edges, and faces, the mesh is often used in computer graphics processing, such as texture mapping. Light field, the representation method closest to the ideal hologram so far, records all light rays reaching a specific point from different directions and positions. With 6-DoF visual representation capability, point cloud, mesh, and light field can all serve as 3D representations for holographic videos. 

\begin{figure}[t]
\centering
\includegraphics[width=0.95\columnwidth]
{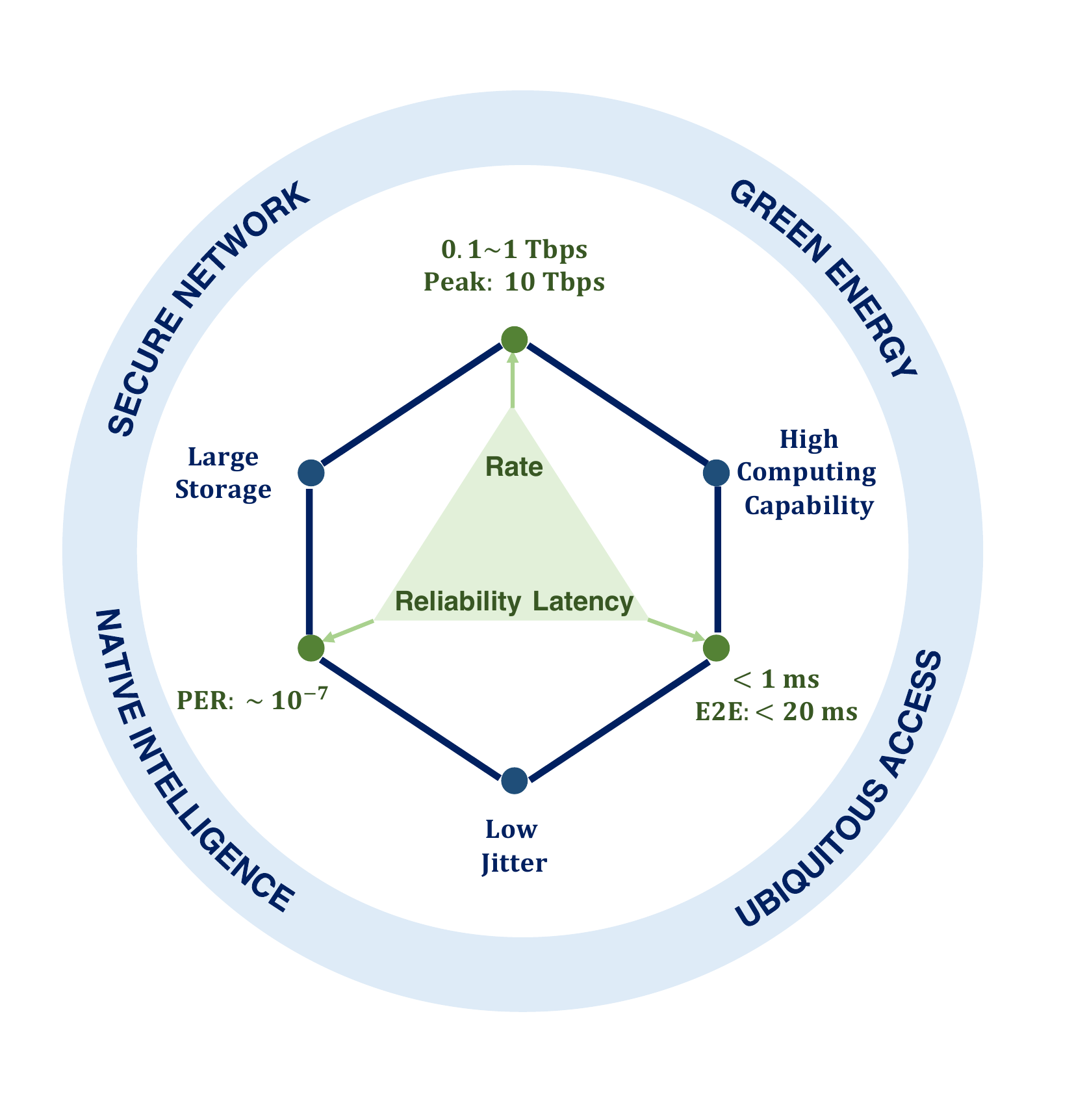}
\caption{Requirements of holographic video communication systems.}
\label{requirements}
\end{figure}

Based on the aforementioned holographic video representations, namely point cloud, mesh, and light field, a general HVC system can be developed, primarily consisting of the following components: acquisition, encoding, transmission, decoding, rendering, and display. The transmission part, which is the key module for elevating the local display of holographic video to HVC, brings about more application scenarios while also posing higher demands on the communication system. To support high-quality HVC services, the requirements of communication systems are broken down into several parts, as shown in Fig. \ref{requirements}. The three core key indicators are transmission rate, transmission latency, and transmission reliability. Specifically, the transmission rate should reach $0.1\sim1$ Tbps, with a peak rate of $10$ Tbps; the air interface transmission latency should be less than $1$ ms, and the network end-to-end transmission latency should be less than $20$ ms; the packet error rate (PER) should be at the level of $10^{-7}$. Additionally, to ensure synchronization between multiple data streams, as well as efficient storage and algorithm processing, the communication system also needs to provide low jitter, high computing capability, and large storage. Furthermore, a large amount of user data will be sent into the network in HVCs. AI models will be applied in various modules of a HVC system. Multi-user support and sustainability are also practical requirements. Thus, secure network, native intelligence, ubiquitous access, and green energy are also essential for the future development of HVCs.

Since there have been significant differences in the methods for processing various types of 3D data so far, to enable more targeted designs and facilitate the implementation of a prototype system, we select a specific data representation suitable for current HVCs. The Gbps-level transmission bandwidth requirement of point cloud data representation is lower than the Tbps-level requirements of light field transmission and generally less than that of mesh representation under the same scale. The point cloud has a regular structure and strong scalability, making it easier to perform pre-processing and post-processing at both the transmitting and receiving ends. For instance, point clouds can be voxelized into voxels before encoding or converted to mesh before texture rendering. However mesh representation faces certain difficulties in rendering non-manifold objects and light field representation is limited by current physical devices, posing challenges for scaling the display. Therefore, point clouds are a better data representation for HVCs at this stage.

Since 2014, the Moving Picture Experts Group (MPEG) has been working on standardizing Point Cloud Compression (PCC), developing two major branches: Video-based Point Cloud Compression (V-PCC) and Geometry-based Point Cloud Compression (G-PCC). DL methods have also been applied to point cloud compression, where the latent features extracted by the autoencoder are compressed by arithmetic coding with the assistance of a hyperprior \cite{wang2021lossy}. These compression algorithms can significantly reduce the data volume for holographic video transmission and provide insights for applying semantic encoding technologies in HVCs.

\subsection{Semantic Communications}
As mentioned earlier, semantic communication is suitable for enabling HVCs, as it can effectively reduce a large amount of redundancy in holographic content. Being a communication paradigm that transmits the meaning of information, semantic communication can extract and compress relevant semantics from holographic content through the semantic encoding module based on the user's needs \cite{zhu2022semantic}. Besides effective semantic extraction and compression, it is also necessary to protect semantic information against channel impairments. Therefore, the introduction of a semantic protection module is required \cite{qin2024ai}. Based on these modules, along with the corresponding decoding and reconstruction parts at the receiving side, the core part of semantic communication, the joint semantic-channel coding (JSCC) framework, is constructed. Leveraging the powerful nonlinear mapping and representation capabilities of DL models, all modules in the framework can be implemented using neural networks. Through joint training of the models, effective and reliable joint coding codewords can be obtained. 

In addition to the core JSCC framework, a semantic communication system also includes important modules, such as semantic knowledge base, semantic sampling, semantic reconstruction, and semantic-aware transmission \cite{qin2024ai}. The semantic knowledge base is usually implemented by pre-trained models or knowledge graph technology, primarily to enhance the efficiency of the semantic extraction and compression modules. Semantic sampling, which is the module preceding semantic encoding, can significantly reduce the amount of transmitted data and alleviate the processing burden of semantic encoding. Semantic reconstruction, performed after semantic decoding, completes different levels of reconstruction according to user needs. Semantic-aware transmission serves as an intermediate component coupling semantic information with advanced transmission technologies such as multiple-input multiple-output and orthogonal frequency division multiplexing. The core framework of semantic communication and other important modules mentioned above can enhance the performance of HVCs at various phases. In the next section, semantic-aware holographic communications will be introduced in detail.

\section{Semantic-Aware Holographic Video Communications}
\label{Section III: Semantic-Aware Holographic Video Communications}
In this section, we explain how semantic communication and HVC are integrated, introducing the overall architecture of a semantic-aware holographic communication system. Based on the proposed architecture, we further discuss potential implementation methods for key modules and show our designs in semantic sampling, JSCC, and semantic-aware transmission. Besides, performance metrics are introduced at the end of this section.

\subsection{Architecture of a Semantic-Aware Holographic Video Communication System}
\begin{figure*}[t]
\centering
\includegraphics[width=1.95\columnwidth]
{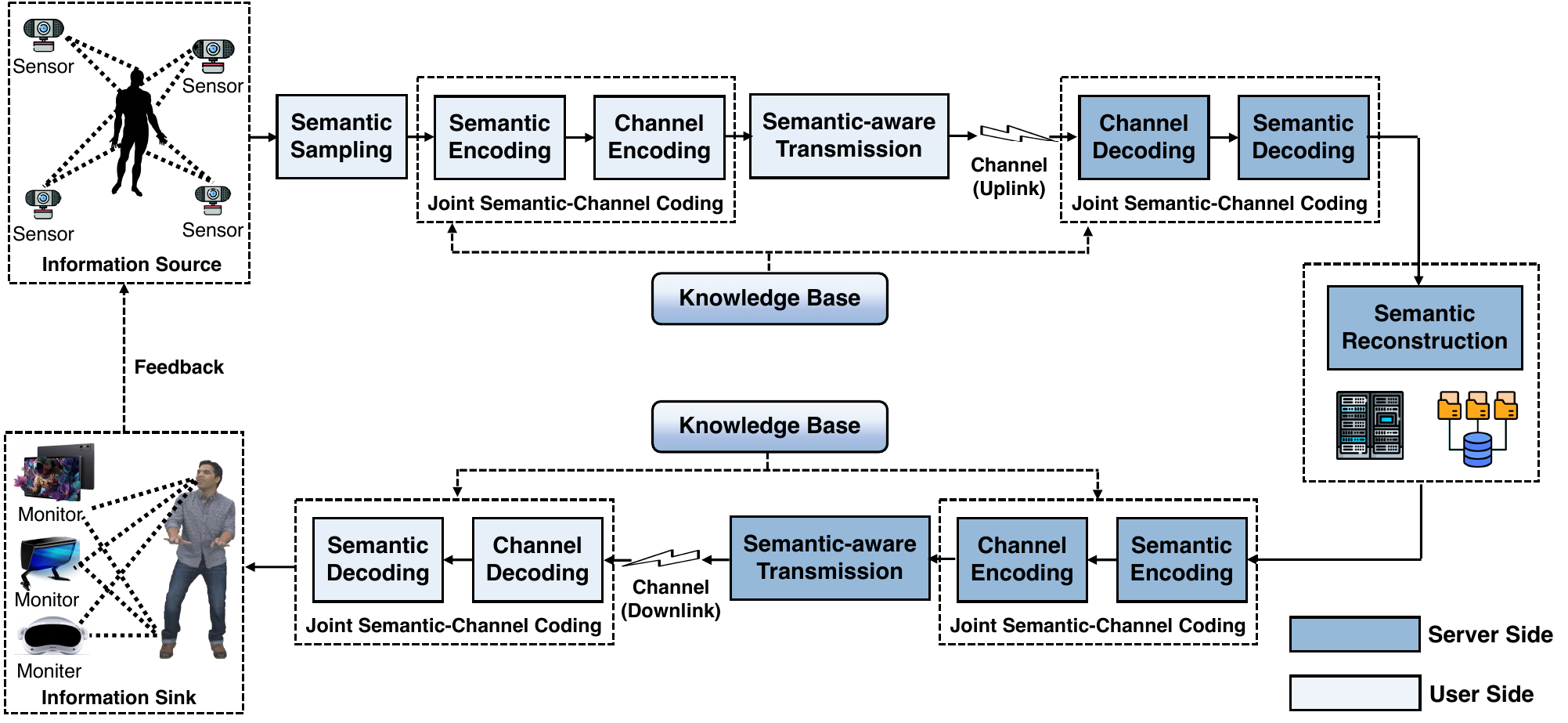}
\caption{The architecture of a semantic-aware holographic video communication system.}
\label{architecture}
\end{figure*}

We propose an architecture for semantic-aware HVC as demonstrated in Fig. \ref{architecture}. This architecture incorporates several advanced semantic communication modules, including semantic sampling, JSCC, semantic transmission, and semantic reconstruction. In addition to tailoring these modules for holographic content, the most distinctive feature of this architecture is the introduction of the server as an intermediate processing node, forming both uplink and downlink transmission links. This design addresses the challenge that, in direct user-user transmission, end devices would be unable to handle the massive requirements for storage and computational resources in HVC.

Specifically, in the uplink, the sensor sends the collected raw holographic data to the semantic sampling module for pre-processing. Sampling parameters, such as the number of samples per unit area and sampling matrix, can be adjusted in semantic sampling module. It can also perform selective quantization or masking operations, such as voxelizing or masking geometric and attribute features of point clouds. The use of sampled data alleviates the processing burden on JSCC. The trained JSCC model outputs informative and robust semantic features for holographic contents. The semantic-aware transmission module will minimize semantic impairments during transmission. After decoding on the server, the server’s powerful computing and storage capabilities enable the timely reconstruction of holographic content of varying quality.

In the downlink, the server can perform targeted transmission based on the quality of holographic content required by the user such as resolution, color depth, and number of viewpoints. The downlink transmission bypasses the semantic sampling and reconstruction modules, directly selecting the appropriate JSCC model for encoding and decoding. This implies that, during the offline phase, the system must prepare various encoding and decoding models of different specifications, or effectively utilize pre-trained knowledge bases to quickly update JSCC models. At the information sink, holographic video can be displayed on different devices such as head-mounted displays or 3D screens. Furthermore, user experience can be fed back to the information source, allowing for adaptive adjustments in sampling and encoding modules.

\subsection{Semantic Sampling and Reconstruction for HVCs}
\label{Section III: B}
Compared to semantic communication systems designed for text, speech, image, and video modalities, semantic sampling, and the corresponding reconstruction model play a more critical role in the semantic-aware holographic video communication system. This is because semantic sampling determines the total amount of content that the subsequent system needs to process. Holographic data still imposes a significant burden on existing devices. Proper sampling can significantly reduce encoding, decoding, and transmission latency while ensuring the quality of user experience.
\subsubsection{Semantic Sampling}
As mentioned in Section \ref{Section II: Overview of Holographic Video Communications and Semantic Communications}, we considered point cloud sampling methods here. Most of the previous sampling methods in point cloud networks are based on mathematical statistics, such as farthest point sampling, Poisson disk sampling, and random sampling. The most classical farthest point sampling method was proposed in PointNet++\cite{qi2017pointnet++}, which is widely used in the sampling step of subsequent point cloud related work. However, in the case of low sampling ratio, mathematical statistics-based sampling methods make the downstream task metrics drop drastically.

As aforementioned, semantic enabled methods are promising to perform more effective point sampling. Therefore, we propose a DL-based sampling model that enables the utilization of semantic information. In the proposed model, a multi-layer perceptron (MLP) embeds points into the latent space first. Then the entire point cloud is divided into patches, each of which consists of a centroid point and its $k$ nearest neighbors. The local attention layer \cite{Zhao_2021_ICCV} processes embeddings in each patch to generate intermediate features, as well as a semantic map. Similar to how pixels in key regions, such as edges, exhibit greater variations in an image, points carrying more significant semantic information tend to show higher normalized standard deviations. Accordingly, we design a down-sampling layer that computes a score for each point feature based on its normalized standard deviation and selects the top M points. These sampled points achieve superior preservation of geometric structures in critical regions and exhibit enhanced task-specific representation capabilities. To perform progressive down-sampling, the local attention and down-sampling layers are applied iteratively.

As for the specific task, we adopt a two-stage training strategy. In the first stage, we train the proposed sampling model and downstream task network with complete point clouds, which means the sampling ratio is $1$. In the second stage, the downstream task network is frozen, and the sampling ratios are randomly selected from $[0,1]$ to train a semantic enabled sampling model. The loss function comprises a reconstruction metric (e.g., Chamfer distance) and a task loss (e.g., cross-entropy loss for classification tasks).

\subsubsection{Semantic Reconstruction}
Point cloud reconstruction refers to the process of recovering the original point cloud after down-sampling. Rather than emphasizing the one-to-one correspondence of the reconstructed points, we focus on ensuring that the reconstructed points form recognizable and meaningful structures. Since the down-sampled data exhibit a markedly different distribution from the original point cloud, reconstructing the complete point cloud without semantic priors becomes highly challenging. To address this issue, a reconstruction network guided by semantic maps can be designed. Specifically, the transmitter and receiver will share the semantic map from semantic sampling. Priors from the semantic map are processed and fused with the sampled features through a skip connection. The reconstructed points are recovered by decoding the fused features through an architecture symmetric to the sampling module.

\subsection{Semantic Coding and Transmission for HVCs} 
\label{Section III: C}
When designing semantic communication systems, JSCC and semantic-aware transmission are closely intertwined. Meticulous design is required to effectively protect the semantics of JSCC encoded outputs during quantization, modulation, precoding, and other transmission processes. It is also important to provide JSCC codewords with different semantic importance for semantic-aware transmission. Compared to the traditional information modalities, holographic video has larger data volumes and richer semantic information. Therefore, for holographic video, the JSCC and semantic-aware transmission modules have greater design space and a more significant impact on the end-to-end performance.

\subsubsection{Deep Learning Based JSCC}
Based on the analysis in Section \ref{Section II: Overview of Holographic Video Communications and Semantic Communications}, point clouds are currently the most suitable 3D representation for HVC. Therefore, the following design and discussion will focus on point clouds.

In computer vision society, pretext tasks and powerful pre-trained models are designed to extract meaningful features of point clouds for various downstream tasks like semantic segmentation and object recognition. Inspired by this approach to semantic feature extraction, we design a JSCC model by integrating the powerful point cloud feature extraction networks PointNet++ and Point Transformer.

In the encoder, PointNet++ \cite{qi2017pointnet++} is employed to perform the initial processing of the input point cloud. By aggregating local neighborhood features, it reduces the number of points while increasing the feature dimension. The preprocessed point cloud is then passed through two parallel semantic feature refinement modules based on Point Transformer \cite{Zhao_2021_ICCV}. In the main branch, two Point Transformer Blocks are used to capture more fine-grained structural features of the point cloud, while in the auxiliary branch, a single Point Transformer Block is used to extract coarse-grained semantic features.

In the decoder, Point Transformer is further applied to refine the noisy features. Then, upsampling is performed to reconstruct the input point cloud. In the upsampling process, the additional points in a neighborhood are generated based on a central point and various coordinate differences. Coordinate differences are learned from the aggregated feature of this neighborhood.

Overall, this JSCC design adopts an asymmetric autoencoder structure and utilizes powerful point feature extraction models. Due to computational and model complexity, this method is currently more suited for handling small-scale point clouds. However, it is capable of maintaining good performance even under extreme channel conditions due to its ability to extract high-quality semantics.

Developing JSCC based on DL-enabled point cloud compression methods by inheriting the autoencoder structure from entropy model-based compression methods and removing the hyperprior part is also a potential solution for HVCs \cite{wang2021lossy}. Besides, since diffusion models have demonstrated outstanding performance in point cloud tasks \cite{Zheng_2024_CVPR}, employing them as the decoder in JSCC for reconstructing noisy point clouds also holds great potential. A diffusion model can be combined with a vector-quantized semantic encoder. By transmitting only the indices, the transmission cost of HVC can be reduced \cite{fu2023vector}, while at the receiver side, the embeddings corresponding to the indices can be used as conditions to guide the denoising process of the diffusion model.

\subsubsection{Semantic-Aware Transmisson}
Semantic-aware transmission aims to efficiently utilize the codewords output by JSCC, ensuring the most effective transmission while preserving the structure of the extracted semantic information. The key in developing semantic-aware transmission models is to ensure that all involved processes are differentiable during backpropagation, so that the optimal channel symbols and waveforms can be obtained through end-to-end training.

For the semantic-aware transmission part, we design a differentiable modulation model. Combined with the previously designed JSCC model,  it forms a joint semantic-channel coding and modulation scheme. Specifically, the semantic features output by the JSCC encoder are treated as probabilities indicating the positions of modulation constellation points. Sampling based on these probabilities avoids the non-differentiability issue that arises when directly quantizing the JSCC output to constellation points.

In addition, to enable semantic-aware transmission, the JSCC output is used to guide the generation of partition points. Constellation points after the partition point will not be transmitted. To make the model channel-adaptive, we concatenate the channel information with the JSCC output to learn more robust features. During training, the joint semantic-channel coding and modulation scheme samples a channel condition in each iteration and uses both the reconstruction quality of the point cloud and the number of adaptively adjusted transmitted symbols as the loss function, thereby obtaining the final model.

\subsection{Performance Metrics for HVCs}  
Similar to the evaluation of semantic communication systems for image or video transmission, the end-to-end performance assessment of HVCs can be divided into subjective and objective quality assessments. Objective quality evaluation uses specific models and metrics to calculate scores for holographic visual quality, but the results often may not align with users' subjective experiences. Subjective quality evaluation requires extensive experiments, where participants are exposed to predefined visual stimuli and provide feedback on the set questions. In HVCs, subjective quality assessment is particularly important due to the potential effects of 3D visual effects and display devices, such as dizziness and fatigue. A comprehensive evaluation of HVC quality requires a subjective experimental design to assess clarity, smoothness, realism, comfort, etc.

Specifically, for HVC systems based on point clouds, the simplest performance metric calculates the point-to-point mean squared error (MSE), known as D1. Considering the perceptual characteristics of the human visual system, the performance metric that calculates the MSE of projection between points and planes is known as D2. By expressing D1 and D2 in terms of peak signal-to-noise ratio (PSNR), we obtain the corresponding D1 PSNR and D2 PSNR, which is a more common form in performance evaluation.

\section{Use Cases} 
\label{Section IV: Use Cases}
In this section, we will demonstrate two use cases about semantic sampling and joint semantic-channel coding and modulation for point clouds to show the potential of semantic communications in realizing HVCs.

\subsection{Semantic Sampling for Point Cloud}
In this case study, we designed and evaluated a semantic sampling scheme based on technologies proposed in Section \ref{Section III: B}. Inspired by attention mechanisms, we calculate an attention score for each point by measuring its correlation with other points. Then, we select $M$ points from the original $N$ points in descending order of their attention scores to achieve a sampling ratio of $M/N$. After downsampling the points to $M$ through tailored semantic sampling, the downsampled point cloud is fed to the classifier. The entire system, implemented with neural networks, is trained using cross-entropy and chamfer distance as the loss functions. Cross-entropy ensures classification accuracy, while chamfer distance provides better guidance for sampling, making the sampled object as close as possible to the shape of the original object.

\begin{figure}[t]
\centering
\includegraphics[width=0.95\linewidth]{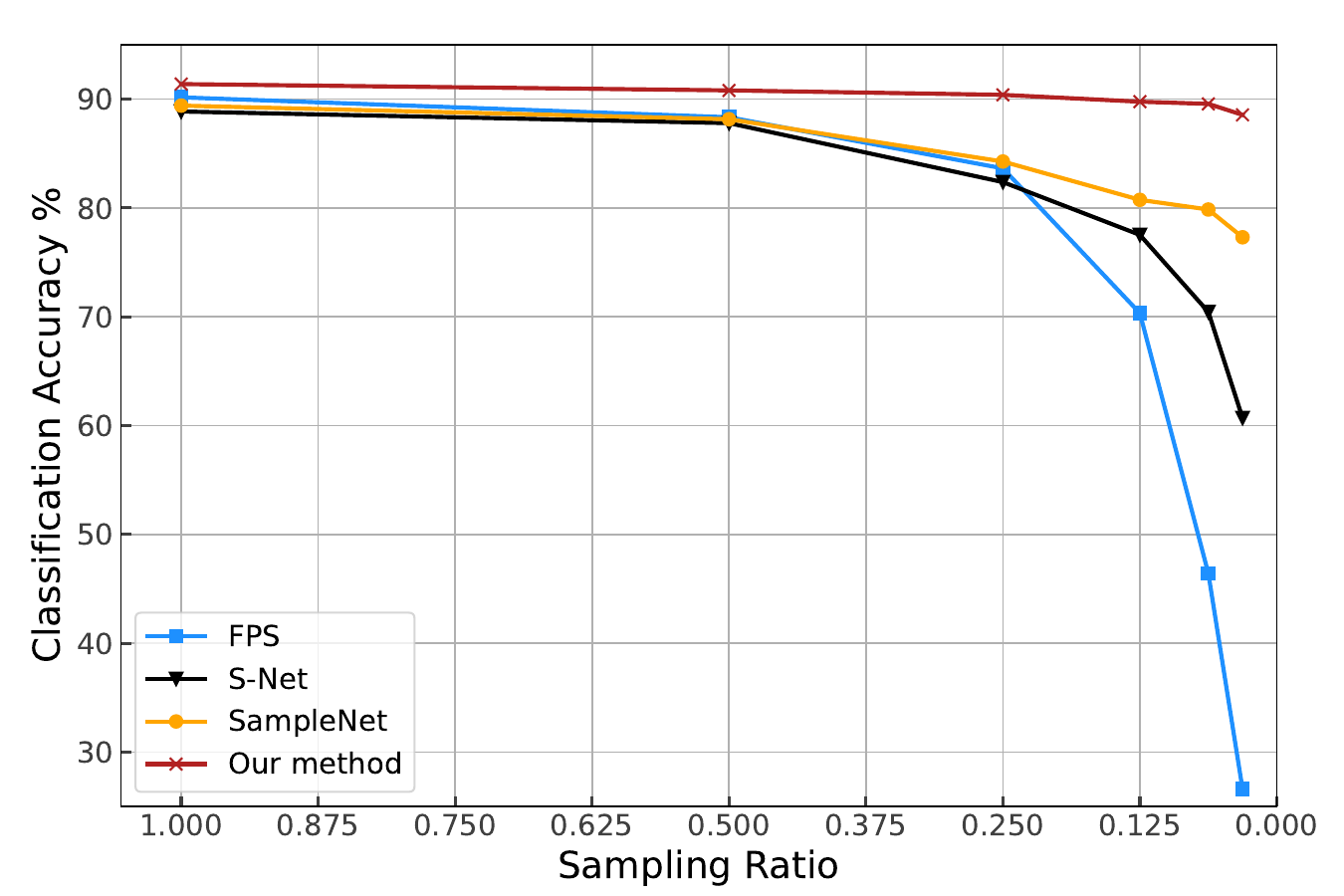}
\caption{Classification accuracy under various sampling ratios.}
\label{ratio_vs_acc}
\end{figure}

\begin{figure}[t]
\centering
\includegraphics[width=0.95\linewidth]{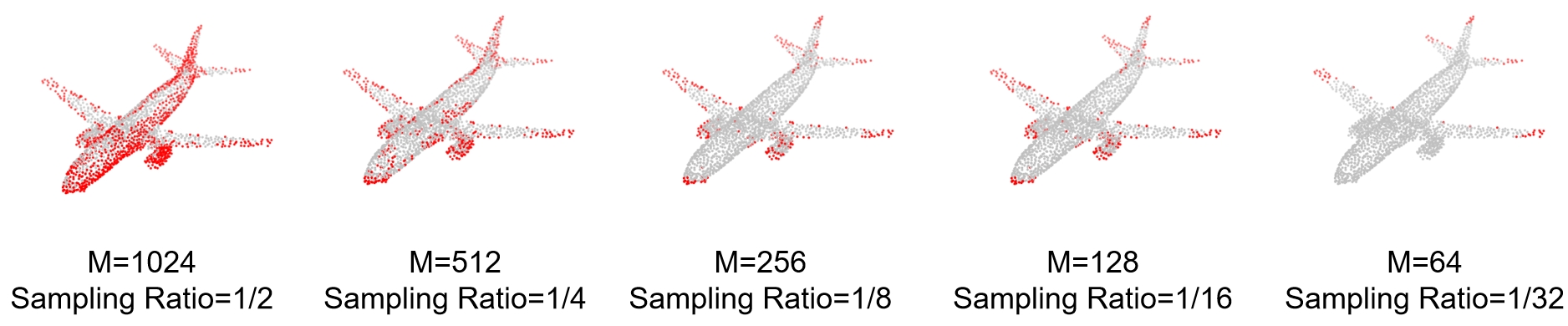}
\caption{Sampled point clouds utilizing our semantic sampling method under various ratios.}
\label{mismatch}
\end{figure}

To evaluate the performance of the proposed scheme, we perform classification task on point clouds with $2048$ points ($N=2048$). As shown in Fig. \ref{ratio_vs_acc}, compared to other sampling methods, such as farthest point sampling (FPS), S-Net \cite{Dovrat_2019_CVPR}, and SampleNet \cite{Lang_2020_CVPR}, our semantic sampling approach has a significant performance improvement, especially in the low sampling ratio regime. This indicates that through employing an attention mechanism, our method captures the semantic information of the point cloud more effectively. FPS, as a traditional sampling method, only ensures a uniform spatial distribution. When the sampling ratio is low, it is difficult for FPS to effectively cover the critical parts of the point cloud. S-Net and SampleNet are DL-based methods to perform non-uniform sampling. However, their performance degradation at low sampling ratios indicates that point cloud semantic extraction based on PointNet++\cite{qi2017pointnet++} can be further improved.

We visualize the downsample point cloud with different numbers of points in Fig. \ref{mismatch}. It can be observed that semantic sampling effectively preserves the structure of the airplane in the example, which also implies that the model optimized for classification accuracy can simultaneously ensure good reconstruction performance.

\subsection{Joint Semantic-Channel Coding and Modulation for Point Cloud}
\begin{figure}[t]
\centering
\includegraphics[width=0.95\columnwidth]
{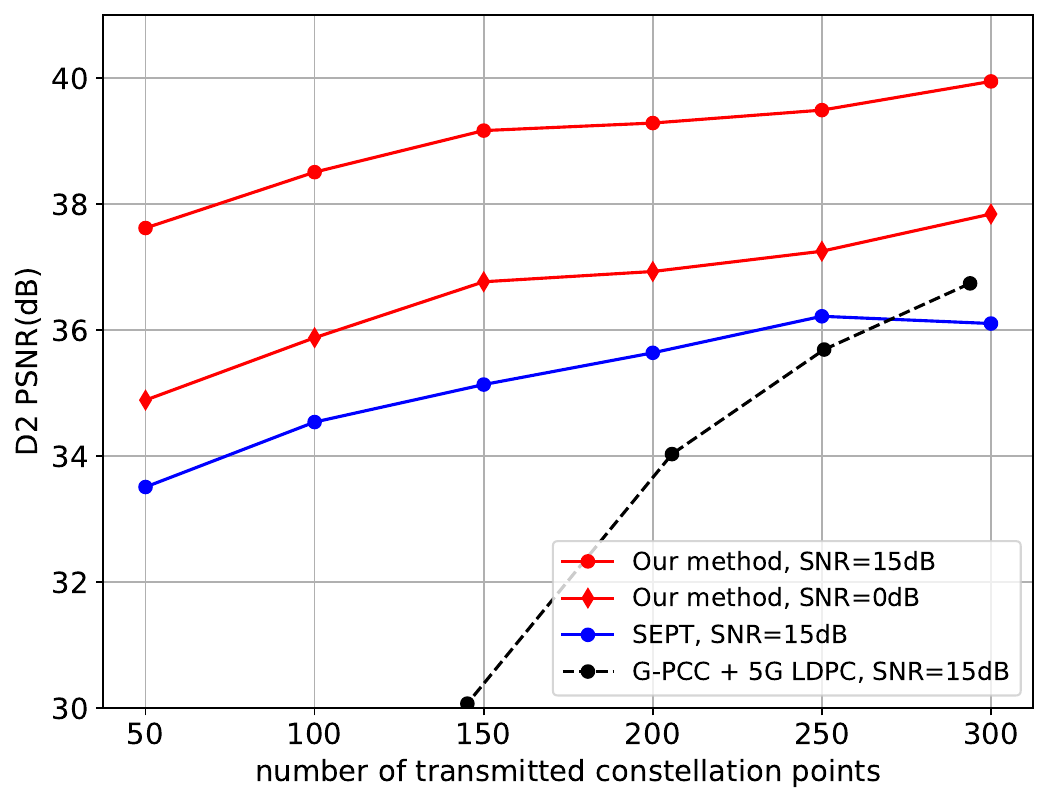}
\caption{The reconstruction performance under Rayleigh fading channels.}
\label{fig6}
\end{figure}

In this case study, we designed and evaluated a joint semantic-channel coding and modulation scheme which covers the JSCC and semantic-aware transmission modules introduced in Section \ref{Section III: C}. We design the JSCC by realizing downsampling modules based on PointNet++ \cite{qi2017pointnet++} and feature refinement based on Point Transformer \cite{Zhao_2021_ICCV} to extract high-quality semantic features. Two JSCC branches with different numbers of feature refinement steps are utilized to generate point semantics with varying granularity. The coupling of JSCC and the standard digital modulation module is then achieved through a probabilistic sampling method.

To verify the performance of the proposed scheme, the traditional separated scheme combined with G-PCC and low-density parity-check codes (LDPC) and the DL-based JSCC scheme for point clouds named SEPT \cite{bian2024wireless} are used as baselines. For the separated scheme, the modulation order and coding rate are selected based on the 5G standard documents, with the optimal modulation and coding scheme (MCS). As shown in Fig. \ref{fig6}, when the signal-to-noise ratio (SNR) is $15$ dB and the number of transmitted constellation points is the same, the proposed method demonstrates an advantage of over $3$ dB in D2 PSNR compared to the baseline. Even at SNR $= 0$ dB, our method outperforms the baseline scheme at SNR $= 15$ dB, since SEPT does not effectively exploit the structured semantics of point clouds. The separate scheme even fails to decode properly due to the cliff effect when the SNR is $0$ dB.

\section{Conclusions and Outlooks}
\label{Section V: Conclusions and Outlooks}
In this article, we investigate 3D data representations and system requirements of HVCs. We introduce semantic communication methods to build efficient and robust HVC systems. Specifically, a semantic-aware HVC architecture is presented, where semantic sampling, JSCC, and semantic-aware transmission modules are designed in detail. Two related use cases are employed to illustrate how semantic communication can enhance the performance of HVCs.

We can further advance the development of semantic communication enabled HVCs in the future by exploring the following topics:

\begin{enumerate}
   \item \textbf{How can the temporal correlation in holographic video be utilized?} Most existing holographic video compression methods focus on intra-frame compression and do not exploit temporal redundancy in 3D space fully. Exploring the semantics of holographic video in the temporal dimension is a promising area for future research in semantic communication for HVC.
   \item \textbf{How to reduce the computational complexity of semantic communication methods?} Although semantic communication technologies for HVCs have demonstrated excellent performance in reducing transmission data volume and improving robustness, current schemes implemented with DL models suffer from very high computational complexity. For modalities such as point clouds, it is highly desirable to design more lightweight attention mechanisms and to strike a balance between model performance and computational complexity.
   \item \textbf{How can semantic communication be used to address light field transmission?} The light field is a data representation closer to ideal holography and it may become one of the data representations for HVC in the future. Effectively converting light fields into representations with more mature processing techniques, such as point clouds or multi-view images, is a promising way to accelerate related research.

\end{enumerate}

\bibliographystyle{IEEEtran}
\bibliography{reference}

\end{document}